# LATE TIME ACCELERATION AND ROLE OF SKEWNESS IN ANISOTROPIC MODELS

S.K.Tripathy

Govt. College of Engineering, Kalahandi, Bandopala, Risigaon, Bhawanipatna, Odisha-766002, India

**Abstract**

We study cosmological models with anisotropy in expansion rates in the context of the recent observations predicting an accelerating universe. In the absence of any anisotropy in the cosmic fluid, it is shown that the role of skewness in directional Hubble rates is crucial in deciding the behavior of the model. We find that incorporation of skewness leads to a more evolving effective equation of state parameter.

## 1. Introduction

Anisotropic cosmological models are studied in recent times. The high resolution Cosmic Microwave Background Radiation (CMB) anisotropy data from Wilkinson Microwave Anisotropy Probe (WMAP) are in good agreement with the prediction of the $\Lambda$ dominated cold dark matter model ($\Lambda$CDM) based upon the spatial isotropy and flatness of the universe[1,2]. However, $\Lambda$CDM encounters some anomalous features at large scale such as (i) observed large scale velocity flows than prediction, (ii) a statistically significant alignment and planarity of the CMB quadrupole and octupole modes and (iii) the observed large scale alignment in the quasar polarization vectors[3]. More precise measurements of WMAP showed the quadrupole $C_2$ and Octupole $C_3$ are usually aligned and are concentrated in a plane about $30^0$ to the galactic plane which suggests that an asymmetric expansion with one direction expanding differently form the other two transverse directions at equatorial plane [4] and signals a non-trivial topology of the large scale geometry of the universe [5, 6]. The statistically significant and planarity of the quadrupole and octupole modes indicates a preferred direction, dubbed as, axis of evil [7]. However, the large scale anomalies in CMB anisotropy are still debatable [8].

While there are many ways to deal with the issue of global anisotropy of the universe, it would be helpful if a simple modification of the FRW model is considered. Recently, some plane symmetric Bianchi-I models or Locally Rotationally Symmetric Bianchi-I (LRSBI) models have been proposed to address the issues related to the smallness in the angular power spectrum of the temperature anisotropy [9-12]. In a recent work, it has been shown that, if the large scale spatial geometry of our universe is plane symmetric with an eccentricity at decoupling of order of $10^{-2}$, the quadrupole amplitude can be reduced drastically without affecting higher multipoles of the angular power spectrum of the temperature anisotropy [9]. Moreover, for a planar symmetry, the universe looks the same from all the points but the points all have a preferred axis. However, it may be noted here that, there still persists uncertainty on these large angle anisotropies and they remain as open problems. LRSBI models are more general than the usual FRW models and are based on exact solutions to the

Einstein Field equations with homogeneous but anisotropic flat spatial sections. LRSBI models have also been studied, in recent times, in different context [13-20].

The observations from distant type Ia Supernovae confirm that presently the universe is undergoing an accelerated phase of expansion [21-23]. The accelerated expansion can be attributed to an exotic form of energy, known as dark energy. A simple candidate for dark energy can be a cosmological constant in the classical FRW model. However, the cosmological constant is entangled with many puzzles like the fine tuning problem, coincidence problem. Some alternative candidates proposed to construct dark energy models are quintessence models [24], phantom models [25], ghost condensate [26] or k-essence [27] and so on. The dark energy provides a negative pressure that generates an antigravity effect driving the acceleration. However, the exact nature of dark energy still remains elusive [28].

In the present work, we have investigated some models with anisotropic expansion rates but isotropic perfect fluid distribution in the frame work of Eintein's general relativity by considering a planar symmetry of the universe. Keeping in view of the present accelerated expansion, we have tried to figure out the role of the skewness in expansion rates in the behavior of the models. The organization of this paper is as follows: In section-2, the dynamics of the LRSBI model is described through the definition of a skewness parameter in the directional Hubble rates and the apparent isotropisation of the model at late times is discussed. In section-3, we have considered widely varying cosmological models predicting late time acceleration to assess the skewness and the directional Hubble parameters. The equation of state parameter defined through the ratio of pressure and energy density $\omega = \frac{p}{\rho}$ is calculated for all the models. At the end, the summary and conclusion of the work is presented in section-4.

## 2. Dynamics of LRSBI model

LRSBI universe is modeled through the spatially *ho*mogeneous and anisotropic metric

$$ds^2 = -dt^2 + A^2 dx^2 + B^2(dy^2 + dz^2) \tag{1}$$

where $A = A(t)$ and $B = B(t)$, t being the cosmic time. The metric corresponds to considering $yz$-plane as the symmetry plane. The eccentricity of such a universe is given by $e = \sqrt{1 - \frac{A^2}{B^2}}$. For a perfect fluid distribution with energy momentum tensor $T^\nu_\mu = (\rho + p)u_\mu u^\nu + pg^\nu_\mu$, the Einstein Field Equations $R^\nu_\mu - \frac{1}{2} R g^\nu_\mu = -T^\nu_\mu$ can be explicitly written in the form

$$H_2^2 + 2H_1 H_2 = \rho \tag{2}$$

$$2\dot{H}_2 + 3H_2^2 = -p \tag{3}$$

$$\dot{H}_1 + \dot{H}_2 + H_1^2 + H_2^2 + H_1 H_2 = -p. \tag{4}$$

Conservation of energy momentum tensor leads to the evolution equation

$$\dot{\rho} + 3H(\rho + p) = 0. \tag{5}$$

$H_1 = \frac{\dot{A}}{A}$ and $H_2 = \frac{\dot{B}}{B}$ are the respective Hubble expansion rates along the symmetry axis and the symmetry plane. An overhead dot over a variable represents ordinary time derivate. In the present work, we have used the unit system such that the velocity of light in vacuum $c = 1 = 8\pi G$.

Let us now define a skewness parameter $\epsilon = H_1 - H_2$ as the difference in the Hubble expansion rates which serves as a measure of departure from the isotropic expansion behavior. $\epsilon = 0$, resembles to an isotropic universe and any other values of it describes an anisotropic one. $\epsilon$ can be related to the shear scalar $\sigma$ as $3\epsilon^2 = \sigma^2$. The mean Hubble parameter or expansion rate can be expressed as $H = \frac{1}{3}(H_1 + 2H_2)$ or $H = \frac{\epsilon}{3} + H_2$. So that the directional Hubble rates are

$$H_1 = H + \frac{2}{3}\epsilon \tag{6}$$

and
$$H_2 = H - \frac{\epsilon}{3}. \tag{7}$$

The Friedman equivalent equations for the anisotropic metric (1) can now be written as

$$3H^2 = \rho + \frac{\epsilon^2}{3}, \tag{8}$$

$$\dot{H} = -\frac{1}{2}(\rho + p) - \frac{\epsilon^2}{3}. \tag{9}$$

It is obvious that, these equations (8)-(9) reduce to the usual Friedman equations for a flat universe in case of vanishing skewness.

The field equations (3) and (4) can nicely be reduced to an evolution equation,

$$\dot{\epsilon} + 3H\epsilon = 0. \tag{10}$$

It is interesting to note that eqn (10) provides the evolution for the skewness parameter just like the evolution of the energy density in a dust filled universe implying that, the skewness in the Hubble rates also evolves through the expansion history. Integration of eqn (10) gives

$$\epsilon = k e^{-\int 3H \, dt}, \tag{11}$$

so that,

$$H_1 = H + \frac{2k}{3} e^{-\int 3H \, dt}, \tag{12}$$

$$H_2 = H - \frac{k}{3} e^{-\int 3H \, dt}, \tag{13}$$

where $k$ is an integration constant and it takes care of the anisotropic behavior of the model. If $k = 0$, then $\epsilon = 0$ and the model reduces to a flat isotropic one. Any other values of $k$ else than zero, provides us an idea about the departure from isotropic case. It is clear from the expression of $\epsilon$ that for a static universe with $H = 0$, $\epsilon$ assumes a constant value otherwise it has to be time dependent. The time varying nature of the skewness parameter $\epsilon$ is decided by the behavior of the Hubble expansion rate.

The equation of state (eos) parameter $\omega = \frac{p}{\rho}$ can now be expressed in terms of the skewness and Hubble rate as

$$\omega = -1 - \frac{2\epsilon^2 + 6\dot{H}}{9H^2 - \epsilon^2}. \qquad (14)$$

The deceleration parameter $q = -1 - \frac{\dot{H}}{H^2}$ can also be rewritten in terms of the skewness as

$$q = \frac{3\epsilon\ddot{\epsilon}}{\dot{\epsilon}^2} - 4. \qquad (15)$$

If we consider that the skewness parameter is proportional to $H_2$ or more specifically $\epsilon = (n-1)H_2$ where $n$ is an arbitrary positive constant, the relative ratio between the Hubble expansion rates along different directions comes out to be a constant quantity i.e.

$$\frac{H_1}{H_2} = 1 + \frac{\epsilon}{H_2} = n. \qquad (16)$$

Such a choice of $\epsilon$ has certain importance in the sense that, it envisages a linear relationship between the Hubble rates in different directions leading to an anisotropic relation between the metric potentials $A$ and $B$ as $A = B^n$. If $n = 1$, the skewness in Hubble rates vanishes and the model reduces to be isotropic. It is a widely prevalent practice to assume such relations to handle anisotropic cosmological models [14, 29-31]. For such a choice of $\epsilon$, the Hubble parameter can be expressed as

$$H = \left(\frac{n+2}{3n-3}\right)\epsilon. \qquad (17)$$

Differentiating the above equation with respect to time and using the fact that $\frac{\dot{\epsilon}}{\epsilon} = -3H$, we can have

$$-\frac{\dot{H}}{H^2} = 3, \qquad (18)$$

which predicts a positive deceleration parameter, $q = 2$. It may be mentioned here that, a positive deceleration parameter signifies a decelerating universe whereas a negative value of $q$ implies an accelerated expansion. In otherwords, the choice $\epsilon \propto H_2$ leads to a constant positive deceleration parameter and it cannot explain the present accelerated expansion of the universe as confirmed in recent observations. This conclusion is valid whether or not we include Newtonian gravitational constant and Cosmological constant with their time variations in the field equations. Even the presence or absence of the cosmological constant in the field equation will result the same [20]. There may be certain deviations from such decelerated behavior of the LRSBI model, if we consider the cosmic fluid to be embedded in magnetic field quantized along the axis of symmetry. Anisotropic fluids with different pressures along different directions where the contributions coming from magnetic field, cosmic strings or domain walls may have a different picture than what has been studied here. However, it is certain that, in an anisotropic LRSBI universe filled with an uncharged perfect fluid with no magnetic field around, the present observations from type Ia Supernovae clearly point towards an evolving relationship between the directional Hubble parameters and hence a time dependent skewness.

The isotropisation in LRSBI model can be assessed. Like most of the Bianchi type models, it is also possible to show that at late time, LRSBI isotropize. Since $H = \frac{\dot{a}}{a}$, where $a = (AB^2)^{\frac{1}{3}}$ is the scale factor, eqn (10) yields,

$$\epsilon = \epsilon_0 \left(\frac{a_0}{a}\right)^3. \tag{19}$$

$\epsilon_0$ and $a_0$ are the values of the skewness parameter and the radius scale factor at the present epoch. In terms of the redshift $z$, we can express $\epsilon$ as

$$\epsilon = \epsilon_0 (1+z)^3, \tag{20}$$

where we have used the fact that $1 + z = \frac{a_0}{a}$. The evolution of $\epsilon$ with the redshift is much clear. At the beginning of the universe, with $z \to \infty$, $\epsilon \to \infty$ and at late time evolution, $z \to -1$, $\epsilon \to 0$. This result indicates that, at the initial epoch, the universe was anisotropic and gradually it evolves to isotropize at late time.

## 3. Cosmologies with accelerated expansion: some examples

It is obvious that, the skewness in Hubble rates $\epsilon$, with the growth of the cosmic time, evolves from a large value to small value at late times. The evolution of $\epsilon$ is closely associated with the behaviour of the Hubble expansion rate. Also it is certain that, a linear relationship between the directional Hubble rates can not predict an accelerating universe. Therefore it is required that we should think for some alternatives that can explain the present scenario. It is necessary that the behavior of the skewness in certain models mimicking an accelerated expansion be assessed. For this purpose we consider certain specific choices of Hubble parameters both from phantom and non phantom regime which are well established and provide the necessary acceleration. The choices are motivated in order to encompass different behaviors of the Hubble rate and scale factor,

Case (i): $H = H_0$=constant, (21)

Case(ii): $H = H_0 e^{\lambda t}$, $H_0$ and $\lambda$ are positive constants, (22)

Case(iii): $H = H_0 - \alpha e^{-\lambda t}$, $H_0$, $\lambda$ and $\alpha$ are positive constants and (23)

Case(iv): $H = \alpha + \frac{\beta}{t^m}$, $\alpha, \beta$ and $m$ are constants. (24)

In Fig.1, we have shown the widely varying behavior of the Hubble rates considered for the present study. Case (i) with a constant Hubble rate gives a constant negative deceleration parameter, whereas case (ii) and case (iii) predict variable deceleration parameters which vary from larger negative values in the past to low negative values in future. At late time case(ii) predicts an uniform expansion. However, for case (iv), there is a transition from deceleration in the past to acceleration at late time of evolution.

**Case (i)**

A constant Hubble rate $H = H_0$ leads to a de Sitter universe with exponential scale factor $a = a_0 \exp[H_0(t - t_0)]$ and it predicts a negative deceleration parameter $q = -1$. The jerk parameter for this model is $r = 2q^2 + q - \frac{\dot{q}}{H} = 1$. The skewness parameter can be obtained as

$$\epsilon = \epsilon_0 \exp[3H_0(t_0 - t)], \quad (25)$$

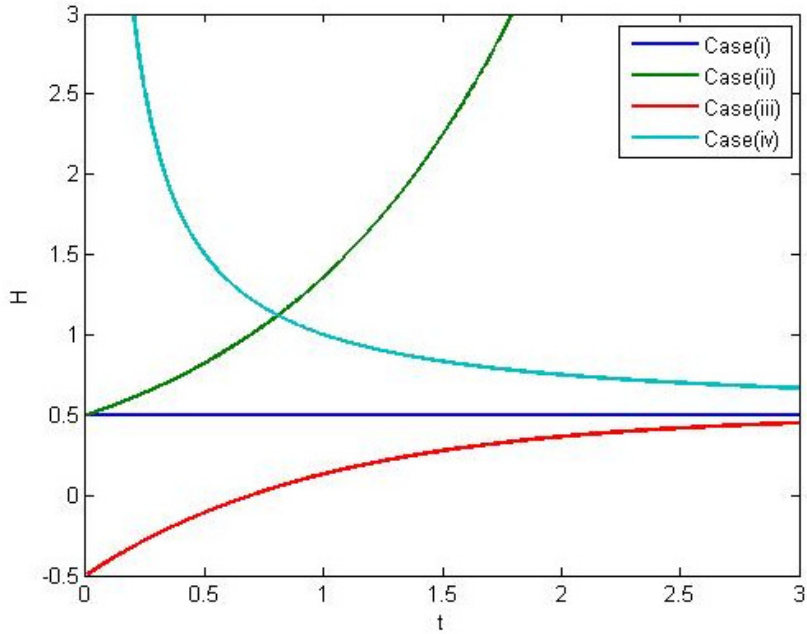

Fig.1. Hubble expansion rates as function of cosmic time.

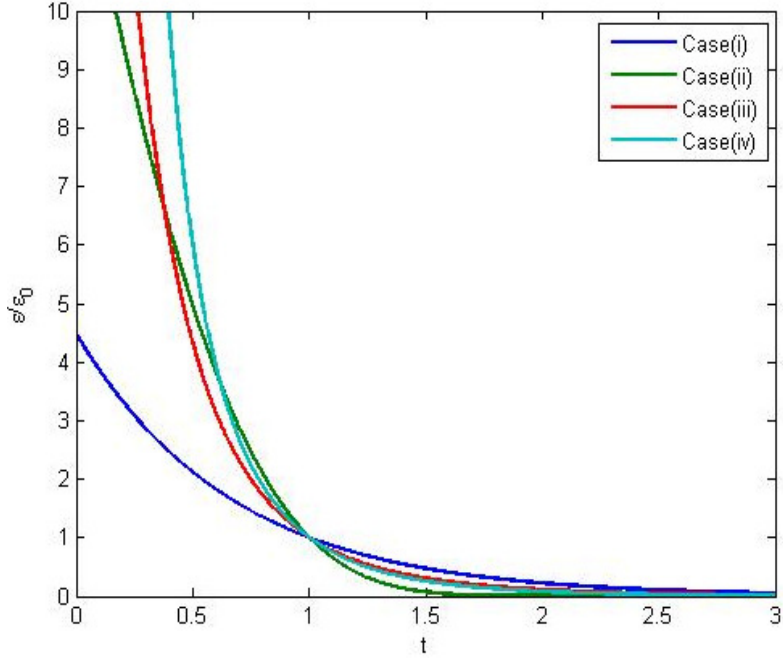

Fig.2. Evolution of the skewness parameter for all the four models considered.

where $t_0$ corresponds to the time at the present epoch. In Fig.2, we have shown the evolution of the skewness in directional Hubble rates. The skewness decreases from $\epsilon = \epsilon_0 e^{3H_0 t}$ in the early time to vanish at late time. Consequently the directional Hubble expansion rates along the axis of symmetry and the symmetry plane are

$$H_1 = H_0 + \frac{2}{3}\epsilon_0 \exp[3H_0(t_0 - t)], \tag{26}$$

$$H_2 = H_0 - \frac{\epsilon_0}{3} \exp[3H_0(t_0 - t)]. \tag{27}$$

It is interesting to note that, even though, the mean Hubble rate is a constant quantity, the directional Hubble rates are time varying. They evolve in quite different manner in the early times. $H_1$ decreases with time whereas $H_2$ increases with time and in late time of evolution they become equal to the mean Hubble rate. In early times, there occurs contraction in one direction ($H_2 < 0$) while there is expansion in other direction ($H_1 > 0$) In the early phase, for a large value of the skewness, i.e. $\epsilon \gg H$, we expect the ratio $\frac{H_1}{H_2} \simeq -2$ as is evident from the Fig.3 (Blue curves). In the figure the upper blue curve describes the evolution for $H_1$ whereas the lower one for $H_2$. At a time when we the skewness is of the order of the mean Hubble rate, $H_1 \simeq \frac{5}{3}H$ and $H_2 \simeq \frac{2}{3}H$, their ratio becomes 2.5.

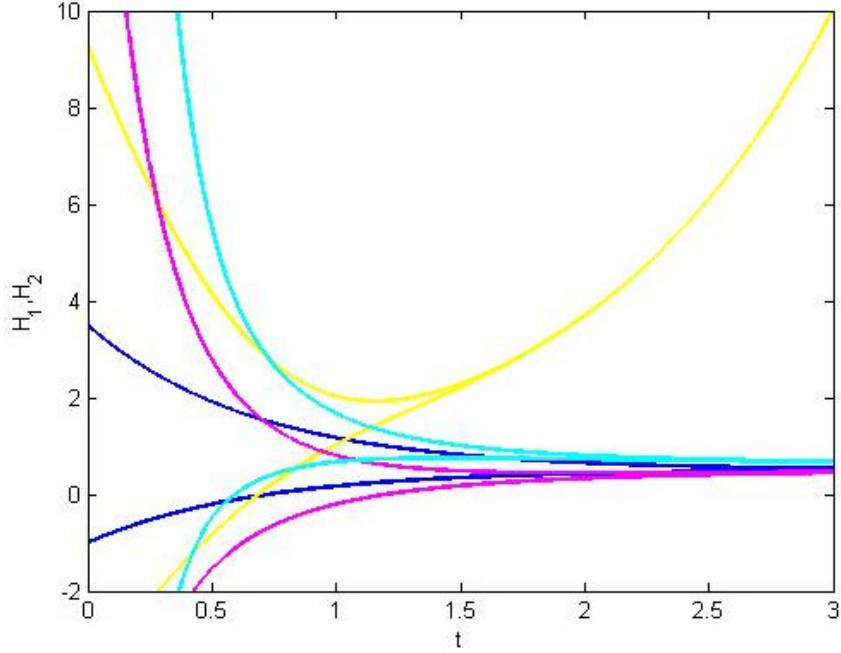

Fig.3. Evolution of the directional Hubble parameters for the four models. The upper curves of a given color are for $H_1$'s and the lower ones for $H_2$'s. The curves in blue, yellow, pink and magenta respectively represent the cases (i), (ii), (iii) and (iv). See text for details.

The equation of state(eos) of the model is obtained as

$$\omega = -1 + \frac{2}{1-9\frac{H_0^2}{\epsilon^2}}, \qquad (28)$$

or in explicit terms,

$$\omega = -1 + \frac{2\epsilon_0^2 \exp[6H_0(t_0-t)]}{\epsilon_0^2 \exp[6H_0(t_0-t)] - 9H_0^2}. \qquad (29)$$

Obviously, the equation of state is not a constant quantity rather it evolves with cosmic time, as shown in Fig.4 (Blue curve), unlike the isotropic case where the eos is a constant with $\omega = -1$. The eos evolves from large negative value to behave as a cosmological constant at late time, when the universe becomes flat and isotropic. However, for infinitely large value of the skewness, the eos becomes positive and in particular for $\epsilon \to \infty$, $\omega \to 1$.

**Case(ii)**

The second case with $H = H_0 e^{\lambda t}$, envisages a phantom picture with little rip behaviour, where Hubble rate tends to infinity in finite future [32-34]. In this model, the Hubble rate is always finite but increases exponentially which generates the strong inertial force. The inertial force in one mass on another mass in the universe, $F_{inertial} = ml(\dot{H} + H^2)$ ( is the mass of the object, $l$ is the separation distance), becomes larger and larger and any bound object is ripped, a phenomenon dubbed as "Little Rip" [32-

36]. In Ref [36], it has been shown that, this asymptotically Little Rip solution is always asymptotically stable. The scale factor for such a model is given by the double exponential expression,

$$a = a_0 \exp\left[\frac{H_0}{\lambda}\left(e^{\lambda t} - e^{\lambda t_0}\right)\right]. \tag{30}$$

The deceleration parameter and the jerk parameter are expressed as

$$q = -1 - \frac{\lambda}{H_0} e^{-\lambda t}, \tag{31}$$

$$r = \frac{\ddot{a}}{aH^3} = -1 - \frac{\lambda}{H_0} e^{-\lambda t} + \left(\frac{\lambda}{H_0}\right)^2 e^{-2\lambda t}. \tag{32}$$

The deceleration parameter and the jerk parameter asymptotically approach to $-1$ at late time. The skewness parameter for this model is given by

$$\epsilon = \epsilon_0 \exp\left[\frac{3H_0}{\lambda}\left(e^{\lambda t_0} - e^{\lambda t}\right)\right]. \tag{33}$$

Consequently, the directional Hubble parameters are

$$H_1 = H_0 e^{\lambda t} + \frac{2}{3}\epsilon_0 \exp\left[\frac{3H_0}{\lambda}\left(e^{\lambda t_0} - e^{\lambda t}\right)\right], \tag{34}$$

$$H_2 = H_0 e^{\lambda t} - \frac{\epsilon_0}{3} \exp\left[\frac{3H_0}{\lambda}\left(e^{\lambda t_0} - e^{\lambda t}\right)\right]. \tag{35}$$

The skewness parameter decrease with the growth of cosmic time and quickly goes to zero as compared to other models (refer to Fig.2). The behavior of the directional Hubble rates are governed by two different factors i.e. mean Hubble rate and the skewness parameter. It is clear from Figs.1 and 2 that, the mean Hubble rate increases with time, on the other hand the skewness in Hubble rates decreases with time. It can be observed from Fig.3 that, in the initial phase, the decrement in skewness affects more to the directional rates. At later times, the increment in mean Hubble rate overshadow the effect of skewness and both the directional Hubble rates behave in the same manner with equal value and behavior with that of the mean Hubble rate.

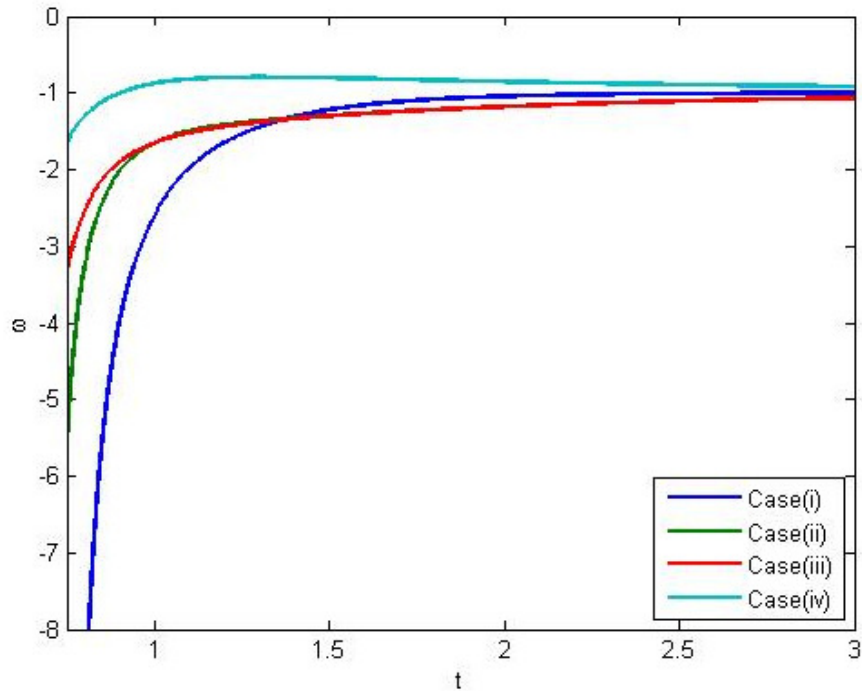

Fig.5. Equation of state parameter as a function of cosmic time.

The eos parameter for the model is

$$\omega = -1 - \frac{\frac{2}{3}\epsilon_0^2 exp\left[\frac{6H_0}{\lambda}(e^{\lambda t_0} - e^{\lambda t})\right] + 2H_0\lambda e^{\lambda t}}{3H_0^2 e^{2\lambda t} - \frac{\epsilon_0^2}{3} exp\left[\frac{6H_0}{\lambda}(e^{\lambda t_0} - e^{\lambda t})\right]}. \tag{36}$$

This expression reduces to $\omega = -1 - \frac{2}{3}\frac{\lambda}{H_0}e^{-\lambda t}$ for the isotropic case. For an isotropic case $\omega < -1$ and $\omega \to -1$ when $t \to +\infty$.

At an early epoch with skewness comparable to the Hubble rate, the eos behaves as

$$\omega = -1.25 - 0.75\frac{\dot{H}}{\epsilon^2} \simeq -1.25 - 0.75\frac{\lambda}{H_0}e^{-\lambda t}. \tag{37}$$

The phantom picture is clearly visible. The eos parameter evolves from a low value of $\omega < -1$ at an early time to $\omega \simeq -1$ at late time of cosmic evolution. $\dot{H}$ is positive and $\omega$ is always less than $-1$ and the deceleration parameter remains always negative.

**Case(iii)**

Another choice for a phantom behavior is Case (iii) with $H = H_0 - \alpha e^{-\lambda t}$ [32, 36, 37], $H_o > \alpha$. This Hubble rate gives an asymptotic de Sitter Universe for positive $H_0$ and $\lambda$ and corresponds to a scale factor given by

$$a = a_0 exp\left[H_0(t - t_0) + \frac{H_1}{\lambda}(e^{-\lambda t} - e^{-\lambda t_0})\right]. \tag{38}$$

In this model, Little Rip does not occur although the Hubble rate and its first time derivative become larger and larger. The inertial force generated by the expansion of the universe is finite [36]. The deceleration parameter for such an universe is given by $q = -1 - \frac{\alpha\lambda e^{-\lambda t}}{(H_0 - \alpha e^{-\lambda t})^2}$, which increases from a very large negative value to asymptotically become $-1$ at late times.

The skewness parameter and the directional Hubble rates are then expressed as

$$\epsilon = \epsilon_0 exp\left[-3H_0(t - t_0) + \frac{3}{\lambda}\alpha(e^{-\lambda t} - e^{-\lambda t_0})\right], \tag{39}$$

$$H_1 = H_0 - \alpha e^{-\lambda t} + \frac{2}{3}\epsilon_0 exp\left[-3H_0(t - t_0) + \frac{3}{\lambda}\alpha(e^{-\lambda t} - e^{-\lambda t_0})\right], \tag{40}$$

$$H_2 = H_0 - \alpha e^{-\lambda t} - \frac{1}{3}\epsilon_0 exp\left[-3H_0(t-t_0) + \frac{3}{\lambda}\alpha(e^{-\lambda t} - e^{-\lambda t_0})\right]. \tag{41}$$

Like other model, the skewness decreases with time. The behaviour of the directional Hubble parameters are like that of the Case(i) ( Pink curves in Fig.3). Unlike case(ii), they tend to a low values at late time. In the early time, their behaviors are more dominated by the skewness: $H_1$ decreases more steadily than $H_2$ increases.

The eos for this model is given by

$$\omega = -1 - \frac{2\epsilon_0^2 exp\left[6-6H_0(t-t_0) + \frac{6}{\lambda}\alpha(e^{-\lambda t} - e^{-\lambda t_0})\right] + 6\alpha\lambda e^{-\lambda t}}{9(H_0 - \alpha e^{-\lambda t})^2 - \epsilon_0^2 exp\left[-6H_0(t-t_0) + \frac{6}{\lambda}\alpha(e^{-\lambda t} - e^{-\lambda t_0})\right]}. \tag{42}$$

This expression reduces to the isotropic one,

$$\omega = -1 - \frac{2}{3}\frac{H_1 \lambda e^{-\lambda t}}{(H_0 - H_1 e^{-\lambda t})^2} \tag{43}$$

at a later epoch when the skewness is negligibly small. As in the case (ii), for an isotropic case $\omega < -1$ and $\omega \to -1$ when $t \to +\infty$.

In an early epoch when the skewness assumes a large value and become of the order of the Hubble parameter, then $\epsilon^2 \sim H^2$ and the eos becomes $\omega = -1.25 - 0.75 \frac{\alpha \lambda e^{-\lambda t}}{(H_0 - \alpha e^{-\lambda t})^2}$. The eos evolves asymptotically to behave as a cosmological constant. One can notice from the Fig.4 that, the behavior of the evolution of the eos is just like the case (ii). In fact, they assume same values at a later epoch. Regarding the stability of the solution of the form (23) as given in case (iii), it has been shown in Ref. [36] that the solution is stable for $3 > \frac{\lambda}{H_0}$. Even though, the solution can be unstable if $3 < \frac{\lambda}{H_0}$, there are still some possibility that the universe could evolve to the de Sitter space-time.

**Case (iv)**

Let us now consider that , $H = \alpha + \frac{\beta}{t^m}$, which gives a scale factor

$$a = a_0 \, exp\left[\alpha(t - t_0) + \frac{\beta}{1-m}(t^{1-m} - t_0^{1-m})\right], \text{ for } m \neq 1. \tag{44}$$

For $m = 1$, one can get, $a = a_0 \left(\frac{t}{t_0}\right)^\beta exp\{\alpha(t - t_0)\}$. The deceleration parameter is

$$q = -1 + \frac{\beta m}{t^{1-m}(\alpha t^m + \beta)^2}. \tag{45}$$

For $m = 1$, $q = -1 + \frac{\beta}{(\alpha t + \beta)^2}$. With suitable choice of the parameters $\alpha$ and $\beta$, this model clearly indicates a transition from decelerated universe in the past to a late time acceleration. The deceleration parameter starts from a positive value in the decelerating phase of the universe and evolves to acceleration zone with negative values at late time. If we chose $\alpha = \beta = 0.5$, with $m = 1$, then in the present epoch the value of deceleration parameter will be -0.5, which is close to the prediction from experiments. The jerk parameter for model with $m = 1$ is

$$r = 1 + \frac{2\beta - 3\beta(\alpha t + \beta)}{(\alpha t + \beta)^3}.\tag{46}$$

The jerk parameter asymptotically approaches to 1. In the initial epoch its value depends on the choice of the model parameters $\alpha$ and $\beta$.

The skewness parameter is now obtained as

$$\epsilon = \epsilon_0 \exp\left[-3\alpha(t - t_0) - \frac{3\beta}{1-m}(t^{1-m} - t_0^{1-m})\right],\tag{47}$$

and the directional Hubble rates are

$$H_1 = \left(\alpha + \frac{\beta}{t^m}\right) + \frac{2}{3}\epsilon_0 \exp\left[-3\alpha(t - t_0) - \frac{3\beta}{1-m}(t^{1-m} - t_0^{1-m})\right],\tag{48}$$

$$H_2 = \left(\alpha + \frac{\beta}{t^m}\right) - \frac{\epsilon_0}{3} \exp\left[-3\alpha(t - t_0) - \frac{3\beta}{1-m}(t^{1-m} - t_0^{1-m})\right].\tag{49}$$

The eos can now be calculated using eqns (14), (42) and (45),

$$\omega = -1 + \frac{2\epsilon_0^2 \exp\left[-6\alpha(t-t_0) - \frac{6\beta}{1-m}(t^{1-m} - t_0^{1-m})\right] - 6m\beta t^{-(m+1)}}{\epsilon_0^2 \exp\left[-6\alpha(t-t_0) - \frac{6\beta}{1-m}(t^{1-m} - t_0^{1-m})\right] - 9\left(\alpha + \frac{\beta}{t^m}\right)^2}.\tag{50}$$

For $m = 1$, the skewness in Hubble rates is expressed as

$$\epsilon = \epsilon_0 \left(\frac{t}{t_0}\right)^{-3\beta} \exp\{-3\alpha(t - t_0)\}.\tag{51}$$

The directional Hubble rates can be obtained using eqns (6) and (7). Like other models, the skewness decreases with time. The directional Hubble rate along the symmetry axis decreases whereas the directional Hubble rate along the plane increases. Their behaviors are just like the case (iii). For this particular case of the parameter $m$, the eos behaves as $\omega = -1 + \frac{2\epsilon^2 - 6\beta/t^2}{\epsilon^2 - 9(\alpha t + \beta)^2}$. At sufficiently large cosmic time, the eos behaves like a cosmological constant. However, at an early epoch, it behaves like $\omega = -1.25 + 0.75 \frac{m\beta}{t^{1-m}(\alpha t^m + \beta)^2}$ which reduces to $\omega = -1.25 + 0.75 \frac{\beta}{(\alpha t + \beta)^2}$ for $m = 1$. It is obvious that $\omega$ evolves from $-1.25 + \frac{3}{4\beta}$, at a cosmic time $t = 0$, with $m = 1$. In the intermediate time it goes above the phantom divide and again approaches to $-1$ at late time.

## 4. Summary and Conclusion

In general, the field equations in anisotropic models are more complicated than isotropic flat FRW models. In order to handle those equations, it becomes necessary to consider some additional physically plausible conditions. In that context, it is a common practice to consider a linear relationship between the directional Hubble parameters which leads to an anisotropic relation between the metric potentials. In plane symmetric models like LRSBI, such an assumption leads to a decelerating universe, what ever may be the matter field taken. However, the conclusion may differ, if we consider a magnetized anisotropic fluid. In view of the recent observations predicting an accelerating universe, it is necessary that to think of alternative relations between the directional Hubble parameters. In the present work, we have defined skewness in the Hubble rates along the symmetry axis and the symmetry plane and derived an evolution equation for it, which behaves just as the energy density in a dust filled universe. In order to assess the behavior of the skewness and directional Hubble rates, we consider four different cosmological models mimicking the accelerated expansion. The mean Hubble rates and corresponding deceleration parameters for the four models behave in widely different manner. However, the skewness in directional Hubble rates and the effective equation of state parameter for all models behave almost in same way. For all the models considered in the present work, the skewness in general decreases with time from large values at the initial period to vanish at late time giving rise to an isotropic universe. The behavior of the directional Hubble parameters depends upon the behavior of the Hubble rate and the skewness. The inclusion of anisotropy in the model makes the equation of state parameter more an evolving one. Even for a model with constant mean Hubble rate, the directional Hubble rates and the equation of state evolve with time. For other models, with already evolving equation of state in isotropic case, the rate of increment or decrement is quite different and is affected by the presence of the skewness parameter. It is observed that, in early time, there occurs contraction in one direction while there is expansion in other direction. The inclusion of anisotropy or the skewness brings about an interesting similarity in the behavior of the equation of state even if the models behave quite differently for their Hubble rates or deceleration parameter.

## Acknowledgement

I am grateful to Prof. D. Majumdar for his help and discussion related to the present work. I gratefully acknowledge the support from Science Academies (Indian Academy of Sciences, Indian National Science Academy and the National Academy of Sciences, India) in form of Summer Research Fellowship to work in Saha Institute of Nuclear Physics, Kolkata, India.